\documentclass{IEEEtran}
\usepackage{cite}
\usepackage{graphics,graphicx,amssymb,amsmath,amsfonts,verbatim}
\usepackage{textcomp,nicefrac}
\usepackage{epsfig}
\usepackage{subfigure}
\usepackage[subnum]{cases}
\usepackage{color}
\usepackage[switch]{lineno}
\usepackage{epstopdf}
\usepackage{bm}
\usepackage{xurl}
\usepackage{listings}

\newcommand\dash{\nobreakdash-\hspace{0pt}}

\def\BibTeX{{\rm B\kern-.05em{\sc i\kern-.025em b}\kern-.08em T\kern-.1667em\lower.7ex\hbox{E}\kern-.125emX}}

\usepackage{enumitem}
\setlist{leftmargin=*}

\begin{document}

\bstctlcite{IEEEexample:BSTcontrol}

\title{Repurposing acquisition devices into trigger-based timing synchronization of breakdown events during MITICA high voltage holding experiments}

\author{\IEEEauthorblockN{A.~Rigoni Garola, L.~Lotto, G.~Manduchi, T.~Patton, R.~Ghiraldelli, C.~Taliercio, M.~Boldrin, L.~Trevisan, P.~Cinetto, P.~Barbato, G.~Chitarin}

\thanks{
Andrea~Rigoni~Garola (andrea.rigoni@igi.cnr.it), T.~Patton, C.~Taliercio, M.~Boldrin, L.~Trevisan, P. Cinetto, and P.~Barbato are with \textit{Consorzio RFX (CNR, ENEA, INFN, Universit\`a Degli Studi di Padova, Acciaierie Venete SpA)}, Corso Stati Uniti 4, Padova 35127, Italy. -- 
L.~Lotto and G.~Chitarin are with \textit{Centro Ricerche Fusione - Universit\`a Degli Studi di Padova} Corso Stati Uniti 4, Padova 35127, Italy. -- 
R.~Ghirardelli is with \textit{Istituto per la Scienza e la Tecnologia dei Plasmi}  Corso Stati Uniti 4, Padova 35127, Italy.
\newline
This work has been carried out within the framework of the ITER-RFX Neutral Beam Testing Facility (NBTF) Agreement and has received funding from the ITER Organization. The views and opinions expressed herein do not necessarily reflect those of the ITER Organization.
This work has been carried out within the framework of the EUROfusion Consortium, funded by the European Union via the Euratom Research and Training Programme (Grant Agreement No 101052200 — EUROfusion). Views and opinions expressed are however those of the author(s) only and do not necessarily reflect those of the European Union or the European Commission. Neither the European Union nor the European Commission can be held responsible for them. \newline
\newline Preprint: Oct 2026 submitted to "MDPI - Sensors"
}
}

\maketitle

\begin{abstract}
%
%

A critical requirement for MITICA — a full-scale prototype of the heating Neutral Beam Injectors hosted at the Consorzio RFX Neutral Beam Test Facility for the ITER experiment — is the capability to withstand a continuous voltage of 1MV across the vacuum gaps insulating the beam source from the grounded vessel. To validate such feature, a dedicated voltage-holding test campaign was conducted throughout 2024 and 2025 using a full-scale mock-up of the beam source. The tests also involved an accurate characterization of the associated breakdown events: vacuum dielectric failures which result in rapid potential drops and generate strong current discharges.

This contribution will present a relative time reconstruction architecture based on cost-effective, embedded RedPitaya (Zynq-7000 FPGA) devices repurposed as timing hubs. These nodes function as configurable trigger multiplexers while simultaneously recording trigger signals as transients to facilitate the offline reconstruction of event sequences. The method allows self-calibration through measuring the static intrinsic delays of the optical fibers and internal logics, generating delay offsets to synchronize acquired waveforms across a sparse, connected-graph topology of both acquisition devices and hubs themselves.

\end{abstract}


\runningpagewiselinenumbers

\section{Introduction}
\label{sec:introduction}

%
%

MITICA is a full-scale prototype of the heating Neutral Beam Injectors for ITER experiment~\cite{Toigo_2017} and is located at Consorzio RFX in Padua, Italy. Across 2024 and 2025, a dedicated voltage-holding test campaign was carried out to validate both the design and the operating range of the system. In particular, a mock-up of the ion source and acceleration grid was built\cite{10200412} to verify the capability of the system to withstand the -1~MV operating conditions during long pulses.

A secondary goal of these tests was the accurate characterization of breakdown (BD) events, namely transient electrical phenomena that occur when the voltage applied exceeds the dielectric strength of the insulating vacuum gaps between conductors. Such events appear as a very rapid voltage collapse accompanied by a high-frequency current discharge, making precise timing and characterization particularly challenging because of their impulsive and stochastic nature and the harsh electromagnetic environment they generate.
%
To reach the nominal voltage of -1~MV required for the beam acceleration, the mock-up underwent a \textit{conditioning} campaign. 

\begin{figure*}
    \centering
    \includegraphics[width=\textwidth]{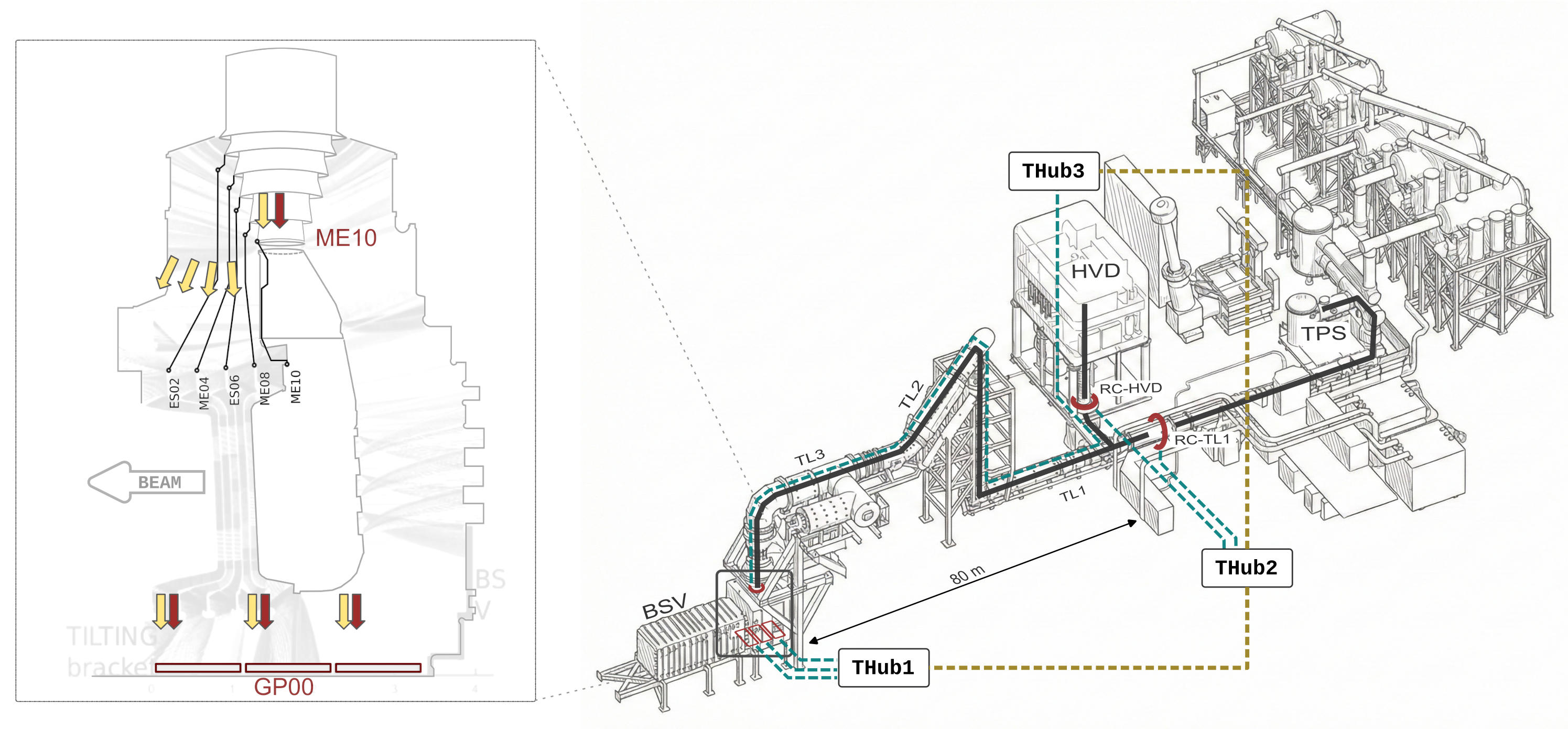}
    \caption{Sketch of the MITICA plant: on the left side a view of the beam source vessel (BSV) showing the source mock-up and the acceleration grid; the marks indicate the current measurement locations where ME10 is the -1~MV potential and GP00 the three floating plates to ground. On the right the components used in the tests with the current path (solid line) and the fiber to propagate trigger (dashed) among the three THubs.  }
    \label{fig:mitica_trigger}
\end{figure*}

The conditioning procedure consists in repeatedly driving the system as close as possible to the BD. This is implemented by a sequence of voltage ramp-ups characterized initially by a set point up to 90\% of the previous BD voltage, followed by a slower rise guided by pressure, microcurrent activity, and X-ray Bremsstrahlung radiation. The aim is to progressively clean the source of impurities and remove surface asperities from the electrodes. As the voltage rises, the system eventually triggers the BD, producing a spontaneous current that extinguishes the arc and resets the potentials. 
A schematic of the plant is shown in Fig.~\ref{fig:mitica_trigger}.
One key requirement for MITICA to operate properly is that all BD events must occur inside the beam source vessel (BSV), in order to avoid damages of the other external structures~\cite{ZANOTTO2023113381}.
In particular, to protect the vessel bushing, the transmission line (TL), the high-voltage deck (HVD, i.e., the room housing the RF generator), and the testing power supplies (TPS), MITICA is equipped with an intrinsic mitigation system composed by vacuum spark gaps on each intermediate voltage stage, together with two core-snubbers on the TL blocking the high-frequency bursts that are reflected back toward the power supplies.

To identify and locate possible failures in the MITICA protection system, during the voltage-holding test campaign, an additional set of diagnostics was installed on the mock-up~\cite{PATTON2023113602}. The objective was to capture the fast current transients flowing through the connections during the BD events. All measurements were time-synchronized, enabling the BD location to be estimated from the relative delays between the signals propagating along the TL and recorded by the different diagnostics.
An internal view of the BSV, shown in Fig.~\ref{fig:mitica_trigger} (left), highlights the current acquisition positions for conductors at different potentials. Red marks indicate the locations for the high-bandwidth BD event measurements and yellow indicate the acquisition of the slow glow-discharge currents observed during conditioning.
Specifically, to capture impulsive arc current events, four fast current transformers (CTs) with a 17~MHz bandwidth were installed inside the vacuum vessel at the -1~MV (ME10) electrode and at three large floor plates (GP00, namely Front, Center, and Rear, in reverse order with respect to the beam injection direction).
To further improve the identification of potential failure sites in the plant, two additional high-bandwidth Rogowski coils were placed on the two main branches of the TL: the first (RC-TL1), located at the midpoint of the Transmission Line component 1 (TL1), measured the current flowing to and from the TPS, while the second (RC-HVD) measured the current flowing to and from the HVD connection.

Electrical modeling indicates that a BD event in MITICA should involve damped current oscillations within a 10~MHz bandwidth and last approximately 100~$\mu$s~\cite{DAN2023113517}. Because of the sporadic nature of such events and the bandwidth of the phenomena, a transient recorder sampling at 125~Msps was selected. In addition, to guarantee the strong electrical insulation required in such an heterogeneous environment, a set of low-form-factor acquisition devices that could be easily insulated and battery-powered was used~\cite{Garola2018AZF}. Four devices were connected to the probes placed in vacuum inside the BSV: three to capture currents flowing through the vessel floor ground plates, and one, sealed inside the -1~MV source, to capture the direct current drained from the main conductor line, as described in~\cite{PATTON2023113602}.

All the devices had to trigger one another in the event of a BD. To this purpose, a common trigger distribution, based on a pre-built optical-fiber multiplexing device, had already been prepared and installed in the plant.
While an NTP-synchronized 1~ksps acquisition was sufficient to detect low-bandwidth phenomena, resolving high-bandwidth BD events required a precise timing reconstruction mechanism, both to accurately assess the relative delays between signals from distant probes and to compensate for time-of-flight (TOF) delays along the connection fibers.
Standard timing solutions based on distributed clock synchronization, such as IEEE-1588~\cite{ieee1588_2008} or White Rabbit~\cite{whiterabbit}, were not easily applicable because of site-specific characteristics and hardware constraints.
In addition, the procurement and deployment of the instrumentation for the test campaign coincided with the post-COVID-19 component shortages, which limited the availability of most commercial solutions.
Since a trigger distribution system to monitor BD events across all devices had already being developed, a custom approach based on that system was eventually adopted. 

Instead of relying on the clocks of the individual devices for time reconstruction, the method reconstructs timing by recording the trigger sequences themselves through a dedicated component called ``Timing Hub'' (THub). The THub, described in Sec.~\ref{sec:methods}, acts both as a trigger dispatcher, distributing triggers across the connected devices, and as a recorder of the relative delays among the triggers generated by those devices.
Since no damaging events happened during the campaign, Sec.~\ref{sec:results} and~\ref{sec:discussion} will describe and discuss the propagation delays of the acquired synchronized signals.

\section{Methods}
\label{sec:methods}

The solution adopted for both slow microcurrent and fast BD current acquisition was to electrically insulate the measuring devices from the vessel potential by using StemLab RedPitaya~\cite{redpitaya_stemlab} boards powered either by batteries or by a DC/DC converters.
The main advantage of such choice is that the measurement front-end can be placed close to the signal source, thereby minimizing cable length while keeping the devices directly referenced to the local ground potential.
This allowed to acquire all measurements with the same architecture, for both microcurrents and BD events, while minimizing the impact of floating grounds and the loop noise pickup.

\subsection{RedPitaya as transient recorder}
\label{sec:redpitaya_rules}

The RedPitaya StemLab 125-14 board integrates an Analog Devices 14-bit LTC2145-14 ADC (formerly Linear Technology) and supports a maximum sampling rate of 125~Msps. The ADC acquisition front-end is directly connected to the Xilinx Zynq-7K on-board FPGA through a parallel bus.
A custom firmware, \texttt{rfx\_stream}, was developed to turn the RedPitaya board into a streaming or transient recorder that can be operated through MDSplus~\cite{Garola2018AZF}.
Here, \textit{streaming} refers to a continuous flow of data acquired and sent to the appropriate consumer during the acquisition, whereas a \textit{transient} is a finite window of data recorded in local memory that, therefore, does not need to be streamed in real time.
In transient mode, the internal FPGA logic acts like an oscilloscope, using the whole internal memory block to acquire the entire buffer at once. The acquisition, then, is suspended and the data is spooled into the internal DDR memory for the host processor. Eventually the device re-arms, ready for the subsequent acquisition.
During the armed state, incoming samples from an AXI-Stream interface are continuously written into the block-RAM-based circular buffer.
When a trigger condition is met---either via an external hardware signal, a software command, or by a threshold comparison (level trigger) on the input channel data---the internal state machine transitions to a checking state to validate the trigger duration.
Upon validation, the logic registers the event, locks the pre-trigger window size, and continues writing post-trigger samples until the configured count is reached.
Finally, the read-control logic transfers the recorded transient window (pre- and post-trigger samples) from the circular buffer to the output AXI-Stream interface and subsequently re-arms the system if multiple triggers are enabled.

The main characteristics of the firmware are that:
\begin{enumerate}
\item the recorder automatically re-arms after each transient;
\item the clock can be considered constant within each transient;
\item the FPGA logic generates a trigger-out signal for every recorded event;
\item the recorder can listen to internal and external triggers simultaneously;
\item the trigger source is recorded for each transient.
\end{enumerate}
These can also be generally regarded as key features for a transient recorder to apply the timing strategy discussed in this work.

\begin{figure}
\centering
\includegraphics[width=0.45\textwidth]{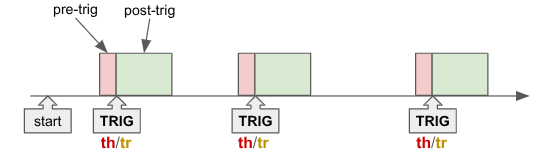}
\caption{Graphical representation of the segments acquired in transient-recorder mode. After being armed (start), the device acquires and stores subsequent events with pre-trigger and post-trigger samples. All triggers are flagged according to whether they originate from other devices or from an internal threshold.} 
\label{fig:prepost_trig}
\end{figure}

\subsection{Repurposing the RedPitaya as asynchronous timing hub}
\label{sec:methods:repurposed}

\begin{figure}
    \centering
    \subfigure[synchronous]{
        \includegraphics[width=0.20\textwidth]{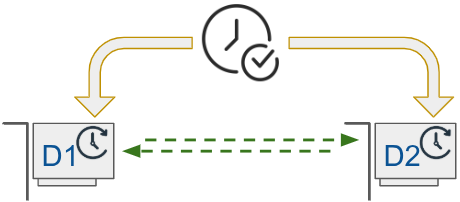}
        \label{fig:sync-async_a}
        }
    \subfigure[asynchronous]{
        \includegraphics[width=0.20\textwidth]{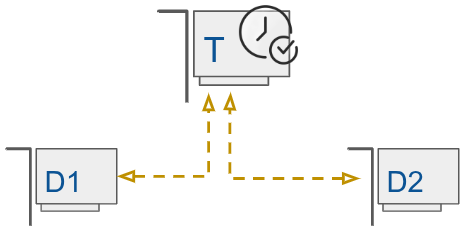}
        \label{fig:sync-async_b}
    }
    \caption{\textit{Synchronous method} (a): all the devices that are triggered, or that collect triggers, share a synchronized clock (either internal or external). Examples include NTP, PTP, and White Rabbit. \textit{Asynchronous} (b): the trigger is collected with respect to a single time reference and subsequently corrected for the signal TOF along the connecting components. The time reference is then kept aligned with absolute time. }
    \label{fig:sync-async}
\end{figure}

Timing reconstruction for short transient events is usually achieved by synchronizing all the devices involved in the trigger sequence, requiring them to share a common time reference to ensure that data is recorded on a unified time-base.
As shown in Fig.~\ref{fig:sync-async_a},
in a fully synchronized architecture, time synchronization protocols establish a common absolute or relative time-base across all the distributed nodes.
This alignment is achieved either by distributing a common physical clock signal or by keeping the internal clock phases locked using digital network delay reconstruction algorithms.
Consequently, each digitizer or local recorder stamps the incoming transient events using its local phase-locked clock.
This direct timestamping enables the immediate correlation of events across separate hardware nodes, and
because all devices are already time-aligned, every event is directly placed within the correct time-base, with no need of any post-processing adjustments.
However, such approach requires specialized, high-precision synchronization hardware, such as IEEE-1588 PTP-compliant PHYs or White Rabbit network, which significantly increases system costs, network complexity, and deployment effort.

To overcome the technical and supply constraints mentioned in Sec.~\ref{sec:introduction}, for the MITICA high voltage tests another method to reconstruct transient timings was considered, exploiting the trigger signals themselves and back-projecting their physical propagation delays. We referred to this approach as ``asynchronous'', as the internal clocks of the participating devices are not aligned to a single master time-base (Fig.\ref{fig:sync-async_b}).
In the asynchronous setup the distributed nodes operate with independent local clocks. Since the clock drift is assumed to be negligible during the transient phase, high-precision synchronization interfaces are unnecessary, allowing the use of a wider range of acquisition devices. Instead of maintaining clock alignment across the entire network, all local trigger-out signals are routed to a centralized timing hub that not only distributes the trigger signal to all the other devices with a minimal delay, but also collects the incoming triggers and records their timing relationship by sampling their logical states as a transient waveform.

\begin{figure}
\includegraphics[width=0.38\textwidth]{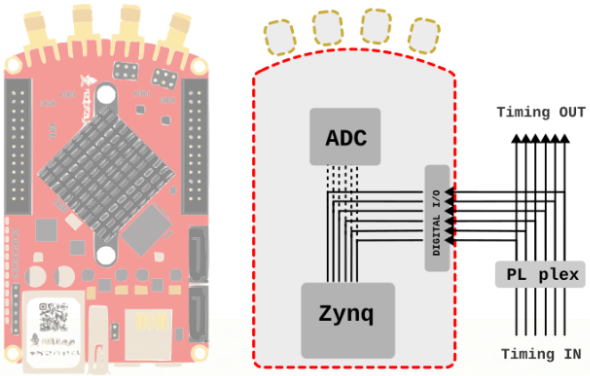}
\caption{Modification of the custom FPGA firmware \texttt{rfx-stream} on the RedPitaya to reroute the ADC bus toward the I/O ports. This makes it possible to interpret the transient signal as the evolution of a bit field representing the state of all trigger lines during the whole BD event.}
\label{fig:rpreroute}
\end{figure}

To this purpose, the idea was to adapt the existing trigger multiplexers connecting them to the available RedPitaya boards in parallel with the internal trigger-dispatch logic, as shown in Fig.~\ref{fig:rpreroute}, thus creating a set of the mentioned ``THub'' components.
For the applied RedPitaya, the same implementation of the \textit{rfx-stream} transient recorder was used, with the activation threshold set to 0x0 (any active bit), in order for the recorded transient samples to directly represent the encoded pattern of the trigger envelope for each channel.
The ports used to acquire the logical values from the ADC parallel bus were directly assigned to six LVDS input ports (TMDS33 signals) through a set of level shifters that converted the logic output of the trigger multiplexer into a differential input signal.
As a consequence, the only required modification was to reroute the bus in the constraints file, thus recompiling the firmware synthesis.

Then a reconstruction algorithm was implemented to convert the bit transitions of the acquired signal -- read sample by sample -- into a measure of the relative delay by counting the number of clock samples between the sequentially detected rising edges of the bits, as summarized in the following Python pseudo-code:
\begin{lstlisting}[language=Python, basicstyle=\ttfamily\scriptsize, breaklines=true, frame=tb]
def getTriggerRelTimes(trigSignal):
  times = np.zeros(NUM_INPUTS, dtype=int)
  mask = 0x1 << PARENT_PORT_ID
  firstId = None
  
  for id in range(len(trigSignal)):
      # Check for new bits
      if (trigSignal[id] & ~mask) != 0:
        # find new flipped channels
        newBits = trigSignal[id] & ~mask
        mask |= newBits
        for idx in range(NUM_INPUTS):
          if newBits & (0x1 << idx): 
            if firstId is None: firstId = id
            times[idx] = (id - firstId) * SAMPLE_NS

   return times
\end{lstlisting}
It is worth noting that, in such a trigger input/output configuration, the only component that must be included in the delay is the trigger-out line, whereas the trigger-in line is used solely to activate the device.

\begin{figure}
    \centering
    \includegraphics[width=0.25\textwidth]{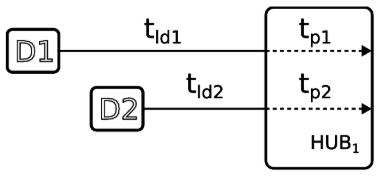}
    \caption{Scheme of the trigger connections for the asynchronous setup.}
    \label{fig:device_connected}
\end{figure}

The THub recorded the relative time among all trigger-out signals emitted by the devices with the precision of the acquisition clock (8~ns for the RedPitaya at 125~Msps). This number did not represent the true event time though. To reconstruct the actual physical sequence of events, the relative arrival time of each trigger had to be corrected by back-projecting the signal using the propagation delays.
In Fig.~\ref{fig:device_connected} the delays are represented as a distance along a line connecting the trigger source to the THub.
Such delays include the trigger time of flight (TOF) through the transmission lines, the electro-optical conversion stages, and the digital processing latencies introduced by intermediate logic both in the device, from the actual trigger event to the time the signal appears at the output port, and in the THub itself from the arrival of the trigger-out signal to the acquisition instant.
The true relative event time $t_{\text{true}, i}$ for channel $i$ was recovered by subtracting the total channel latency offset $t_{\text{dl}, i}$ and the processing delay inside THubs $t_\text{p}$ from the recorded trigger signal time $t_{\text{s}, i}$:
\begin{equation}
\label{eq:back_project}
t_{\text{true}, i} = t_{\text{s}, i} - \left( t_{\text{dl}, i} + t_\text{p} \right)
\end{equation}
While the processing delay $t_\text{p}$ inside the THub can be considered equal for all the input channels, the $t_{\text{dl}, i}$ component is further composed by the propagation delay through the transmission lines and the processing delay inside the source device, that can come either from internal threshold ($\Delta_\text{th}$) or external trigger ($\Delta_\text{tr}$) delay:
\begin{equation}
\label{eq:delay_compoenents}
t_{\text{dl}, i} = l_i \tau_\text{fib} + \left( \Delta_{\text{tr}, i}, \Delta_{\text{th}, i} \right)
\end{equation}

\begin{figure}
    \centering
    \includegraphics[width=0.45\textwidth]{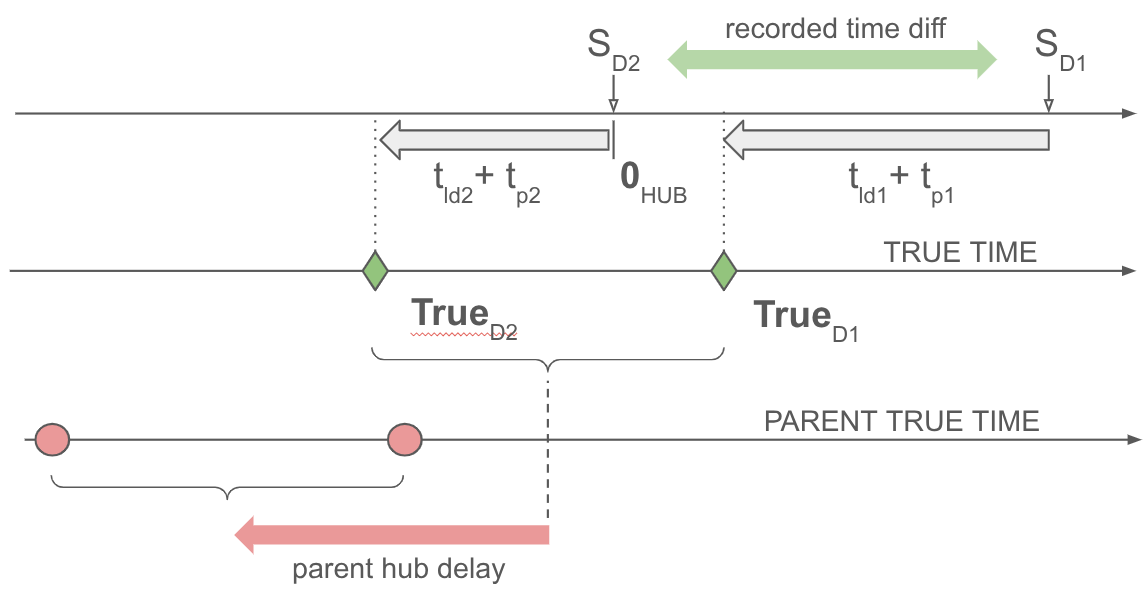}
    \caption{Sketch of relative time reconstruction. The first timeline shows the trigger events as recorded in the trigger hub, while the second timeline shows the same events (green diamonds) after back-projection through their propagation delays. The third plot shows the events (red circles) shifted by the propagation delay of a multi-THub connection.}
    \label{fig:truetime}
\end{figure}
Fig.~\ref{fig:truetime} shows two trigger events from two different devices in the time domain. The first timeline shows the trigger events as recorded by the trigger hub, while the second timeline shows the same events after the back-projection through their propagation delays.
The reconstructed time in this plot has no absolute reference, and the entire timing reconstruction is relative to the selected trigger event. In our implementation, the absolute timing was established by applying the NTP to the trigger-hub signal, since the required precision for correlating BD events with slowly acquired quantities was in the millisecond range. The absolute time base for all slow acquisitions was kept synchronized through an NTP-corrected internal clock using the approach reported in~\cite{manduchi2024mixed}.
In the example, the device labeled D2 is chosen as the actual $t_0$ reference event, which can be assumed to match the NTP time of the hub acquisition plus $S_{D2}$.
It is worth noting that if a more precise absolute correlation is needed, one option is to include a GPS module as one of the connected devices, allowing the trigger to generate a precise absolute timestamp for the entire relative timing chain.

\subsection{Timing distribution}

\begin{figure}[h]
    \centering
    \includegraphics[width=0.25\textwidth]{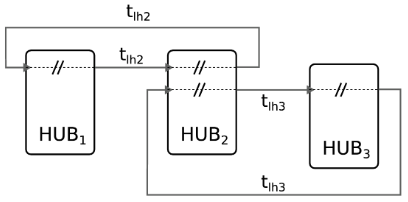}    
    \includegraphics[width=0.20\textwidth]{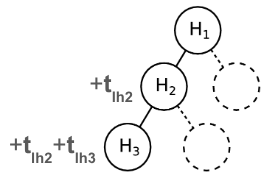}
    \caption{Scheme of multiple THubs connected in a tree topology, where all the child hubs need to be corrected by the sum of the propagation delay path from the root THub.}
    \label{fig:hubs_connected}
\end{figure}

The flexibility of using THubs as intermediate trigger-distribution components lies in the possibility to reconfigure the dispatch mapping, allowing users to enable or disable signal propagation from any given trigger channel to any other. This characteristic is necessary in order to avoid that the devices trigger themselves, but it is also useful for building a trigger-distribution system based on a set of THubs connected in a tree topology. An example is shown in Fig.~\ref{fig:hubs_connected}, where THub1, THub2, and THub3 are connected in a chain.
%
The possibility to connect THubs in a chain or a tree relies on two assumptions:
\begin{enumerate}
\item the THub does not propagate the trigger back to the same port from which it receives it, thus avoiding propagation loops;
\item for the same reason, the recorded time bit field must exclude the parent-THub port when accounting for trigger delays.
\end{enumerate}
If these two conditions are satisfied, a child THub can be treated as just another device connected to the network, and the timing-reconstruction algorithm described above can be used to recover the true time with respect to the parent time base by rigidly shifting all the recorded event times by the propagation delay from the parent to the child. Recursively applied, any connected Thub is eventually corrected by the sum of propagation delays starting from the root, as sketched in Fig.~\ref{fig:hubs_connected}.
This further time adjustment is also shown in Fig.~\ref{fig:truetime}, in the third plot, where the group containing both D1 and D2, considered from the child-THub time base, is shifted by the propagation delay from the parent.

It is worth noting that the hierarchical trigger organization is purely algorithmic, meaning that the topology can be configured at any time, also after the acquisition.

\subsection{Delay calibration}
\label{sec:delay_calibration}

The key assumption behind the asynchronous trigger-timing reconstruction is that the delay associated with trigger propagation among the devices remains constant over time and can therefore be measured. In Fig.~\ref{fig:device_connected}, the total delay is divided into two main components: $t_{ld}$, which mainly represents the propagation delay introduced by the transmission lines together with the logical processing of the trigger signal in the sending device, and $t_{p}$, which represents the internal THub propagation time.
While the internal THub delay is fairly stable and can be considered constant, the delay introduced by the transmission lines depends on the total cable length and on the delay added by the acquisition device.
The TOF in the trigger connections can be assumed mainly proportional to the fiber length, whereas the internal propagation time depends on the acquisition parameters:
\begin{enumerate}
\item{$\Delta_{tr}$ -- } the time delay of the external trigger propagation, which is almost constant but may depend on any algorithm applied to the trigger signal, such as decoding or trigger conditioning;
\item{$\Delta_{th}$ -- } the time delay associated with the internal signal-threshold trigger, which strongly depends on internal parameters, such as the threshold algorithm and the sampling frequency.
\end{enumerate}

\begin{figure}
    \centering
    \includegraphics[width=0.34\textwidth]{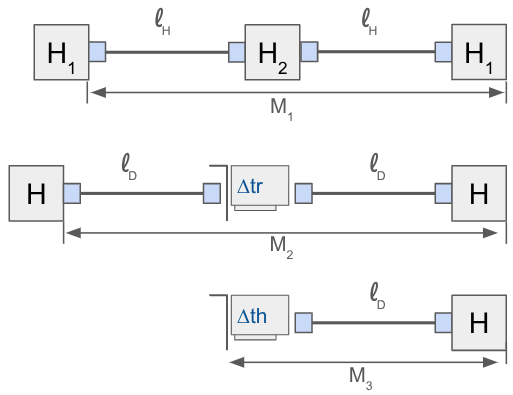}
    \caption{Trigger calibration measurements that need to be performed: $M_1$ is the round trip delay of two connected THubs, $M_2$ is the round trip delay of a externally triggered device, and $M_3$ is the threshold delay propagated through the device fiber. Both $M_1$ and $M_2$ can be in principle self assessed by the hub, while $M_3$ is parameter dependent and need to be provided by the device.}
    \label{fig:trig_measures}
\end{figure}

Since a THub is connected to both ends of the devices through the external trigger propagation, the asynchronous architecture allows the self-calibration for the delay offset introduced by the fiber cables; this can be done by generating a trigger event from the THub itself and by measuring the round-trip delay on the same trigger line. 
Due to the limited implementation time available for the MITICA tests, instead of exploiting the self calibration mechanism, we applied the procedure manually, sending a trigger from an external input and measuring the delay recorded by the THub.
Fig.~\ref{fig:trig_measures} shows the two measurements performed to calibrate the delay. In general, by exploiting the linear dependence of the fiber propagation delay, the internal delay can be also estimated as the bias of the calibration relation with respect to the line length.
The resulting values were set to 450~ns for the 100~m fiber connecting THub1 (vessel) to THub2 (transmission line), and 140~ns for the 30~m fibers connecting THub2 to the devices at the Rogowski coils.
The values of $\Delta_\text{tr}$, and $t_p$ were both externally measured and cross validated by the mentioned calibration procedure; they were set to $100 \text{ns}$ and $80 \text{ns}$ respectively. 
The delay introduced by the threshold trigger $\Delta_\text{th}$ is not constant, but depends on the sampling frequency as follows. 
\begin{equation}
    \label{eq:delta_th}
    \Delta_\text{th} = \Delta_\text{d} + \frac{N_\text{th}}{f_s}
\end{equation}
where: $\Delta_\text{d}$ is the device internal activation time, measuring 85~ns for the RedPitaya, $f_s^{-1}$ is the sampling period, and $N_\text{th}$ is the number of samples to activate the threshold. This formula reflects the algorithm adopted, but a general value might depend on the actual implementation of the thresholding mechanism.

\section{Results}
\label{sec:results}

The MITICA high voltage tests were organized into three main campaigns (respectively named B, C and D~\cite{PATTON2023113602}) aimed at optimizing the conditioning procedure and assessing the voltage-holding performance of the beam source. Campaign B used the TPS, limited to 500~kV, and reached a maximum voltage of 321~kV. Following an upgrade of the TPS, Campaign C focused on assessing the maximum achievable voltage in high vacuum reaching a maximum value of 446~kV, with no clear evidence of conditioning voltage saturation. Finally, Campaign D evaluated the performance of an Intermediate Electrostatic Shield (IES) biased at $-600$~kV, which splits the 1~m vacuum gap to improve the insulation.
Across these three campaigns, a substantial dataset of 3100 pulses was collected, capturing a variety of BD current transients and micro-discharge activities under different experimental conditions.
Since problematic ``out-of-vessel'' events did not occur, in this contribution we restrict the analysis to the time delays of the signals recorded at the different acquisition positions.

\begin{figure}[h]
    \centering
    \includegraphics[width=0.48\textwidth]{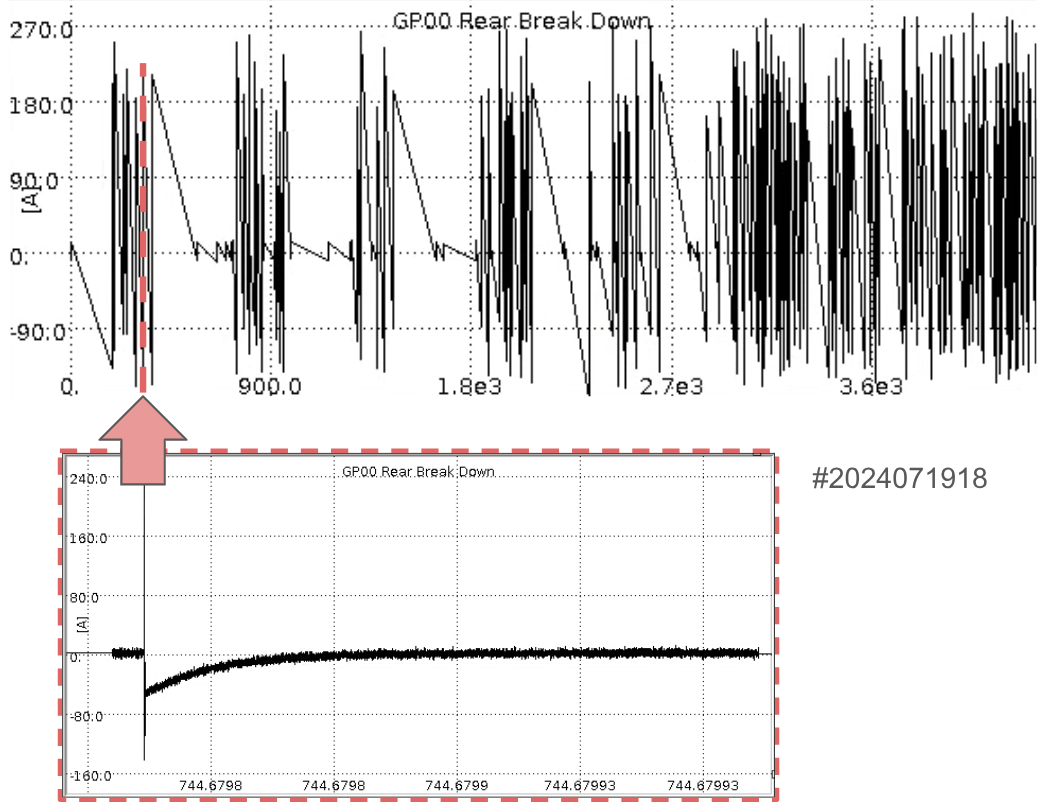}
    \caption{Example of a false BD event recorded by the RedPitaya boards in GP00 (bottom trace), causing many triggered acquisitions (top trace) for all the connected devices.}
    \label{fig:fake_breakdown}
\end{figure}

The digitizer sealed in vacuum and attached to ME10 was damaged by an over-current event and therefore was able to acquire only a very small fraction of the pulses observed during the campaigns. Its recorded values were consequently difficult to interpret and compare with the other sensors and, for this reason, were excluded from the present analysis.
In addition, very early in the campaign an electrical issue related to the grounding of the RedPitaya boards at GP00 became evident. The insulation experienced sporadic electrostatic discharges that generated a large number of false BD events, which were recorded and propagated throughout the connected devices, as shown in Figure~\ref{fig:fake_breakdown}. This produced a large number of spurious archived samples, and may have also masked some real events not recorded because occurring while the RedPitaya was unarmed and storing data.

To clean up the event set, an unsupervised classification approach based on a prepared dataset containing all acquired transients was adopted. This classification helped both to identify signals with similar features, such as sampling frequency and absence of device faults, and to identify and remove false events.

\subsection{Unsupervised Classification of the Breakdown Events}
For the unsupervised feature extraction and classification of the high-dimensional BD signals, a 1D Convolutional Autoencoder (CAE) based on Google Inception-style multi-scale convolutional blocks~\cite{7298594} was used.
The primary goal of this model was to compress the time-series signals into a low-dimensional embedding while preserving critical morphological structures such as sharp spikes and high-variance bursts.

The model architecture consists of a classical symmetric encoder-decoder structure, already tested on other diagnostics~\cite{garola2021diagnostic, orlandi2026data}, but implemented in this application with 1D Inception-style multi-scale convolutional layers.
\begin{figure}
    \centering
    \includegraphics[width=0.48\textwidth]{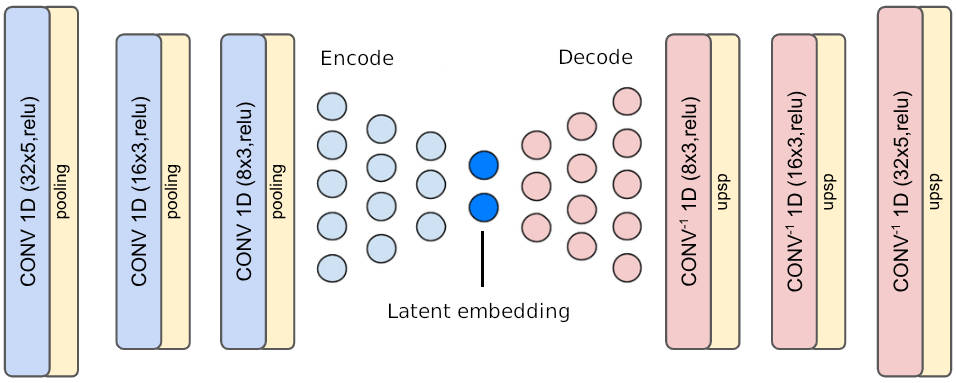}
    \caption{1D convolutional Inception-style autoencoder architecture: three mixed size convolutional layer feed three dense feedforward layers to produce a 32-dim embedding.}
    \label{fig:ae_model}
\end{figure}
These layers apply simultaneously small, medium, and large kernels (sizes 3, 15, and 63) to capture both high-frequency localized features and low-frequency oscillations.
The encoder accepts a resampled input signal of length $N = 4096$, applies three successive 1D blocks with 64, 32, and 16 total filters, respectively, and uses a stride of 2 for downsampling.
The output of the convolutional layers is flattened and passed through a deep dense network with 256, 128, and 64 nodes; finally, a linear dense layer forms a latent bottleneck embedding of dimension $d = 32$.
The decoder approximately reverses this process to reconstruct the original signal.
Each transposed block includes a 1D upsampling layer to restore the signal's spatial resolution while avoiding checkerboard artifacts.

The standard Mean Squared Error (MSE) often fails to accurately reconstruct transient anomalies such as sharp spikes or high-variance bursts in noisy signals. To address this limitation, the model uses a custom loss function, referred to as spike-weighted MSE, that heavily penalizes errors in high-amplitude regions or intervals with high local variance.

The total loss $\mathcal{L}_{\text{total}}$ is defined as:
\begin{equation}
\label{eq:total_loss}
\mathcal{L}_{\text{total}} = \mathcal{L}_{\text{amp}} + \beta \cdot \mathcal{L}_{\text{energy}}
\end{equation}
where $\beta = 5.0$ is a weighting hyperparameter.

The amplitude-weighted loss component $\mathcal{L}_{\text{amp}}$ is given by:
\begin{equation}
\label{eq:amp_loss}
\mathcal{L}_{\text{amp}} = \frac{1}{N} \sum_{i=1}^{N} (y_i - \hat{y}_i)^2 \cdot w_i
\end{equation}
where $y_i$ is the true signal value, $\hat{y}_i$ is the reconstructed signal, and the point-wise weight $w_i$ is defined as:
\begin{equation}
\label{eq:weight_i}
w_i = 1.0 + \alpha \cdot |y_i|
\end{equation}
with $\alpha = 5.0$.

The local burst energy-weighted loss component $\mathcal{L}_{\text{energy}}$ measures the squared difference between the moving average energy of the true and reconstructed signals:
\begin{equation}
\label{eq:energy_loss}
\mathcal{L}_{\text{energy}} = \frac{1}{N} \sum_{i=1}^{N} (E_i - \hat{E}_i)^2
\end{equation}
where the local energy $E_i$ and $\hat{E}_i$ are calculated over a moving window of size $W = 64$ using a 1D convolution:
\begin{equation}
\label{eq:true_energy}
E_i = \frac{1}{W} \sum_{j=i-W/2}^{i+W/2} y_j^2
\end{equation}

To classify the BD signals into distinct physical regimes, a multi-stage clustering pipeline was designed, that combines deep-learning representations with hand-crafted morphological and spectral features.
The preprocessing first involves the resampling and normalization of the signals; then, two complementary sets of features are extracted for each segment.
One is the 32-dimensional latent embedding vector produced by the encoder of the trained AE model; the other is a 44-dimensional vector of extracted signal features, including time-domain statistics, peak counts, signal-envelope properties, zero-crossing rates, burst-duration statistics, energy-distribution metrics, and FFT-based spectral features.

To construct a unified clustering space, the latent embeddings and robustly-scaled hand-crafted features are first individually compressed using PCA.
They are then concatenated, standardized, and further reduced via a second PCA stage to 24 components.
Finally, Uniform Manifold Approximation and Projection (UMAP)~\cite{McInnes2018} projects this combined feature space into an 8-dimensional manifold to facilitate clustering.
In the final stage, the unsupervised clustering groups the signals into four target classes.
For this dataset we evaluated a diverse set of candidate algorithms, including K-Means, Gaussian Mixture Models (GMM), and Agglomerative Clustering, and ultimately selected the GMM as the best option.

\begin{figure}
    \centering
    \subfigure[]{
        \includegraphics[width=0.4\textwidth]{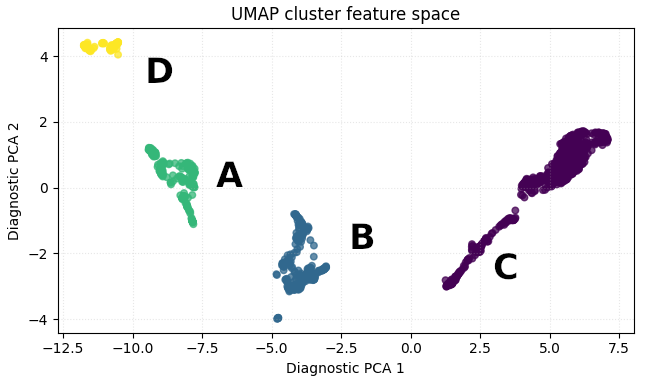}
    }
    \subfigure[]{
        \includegraphics[width=0.4\textwidth]{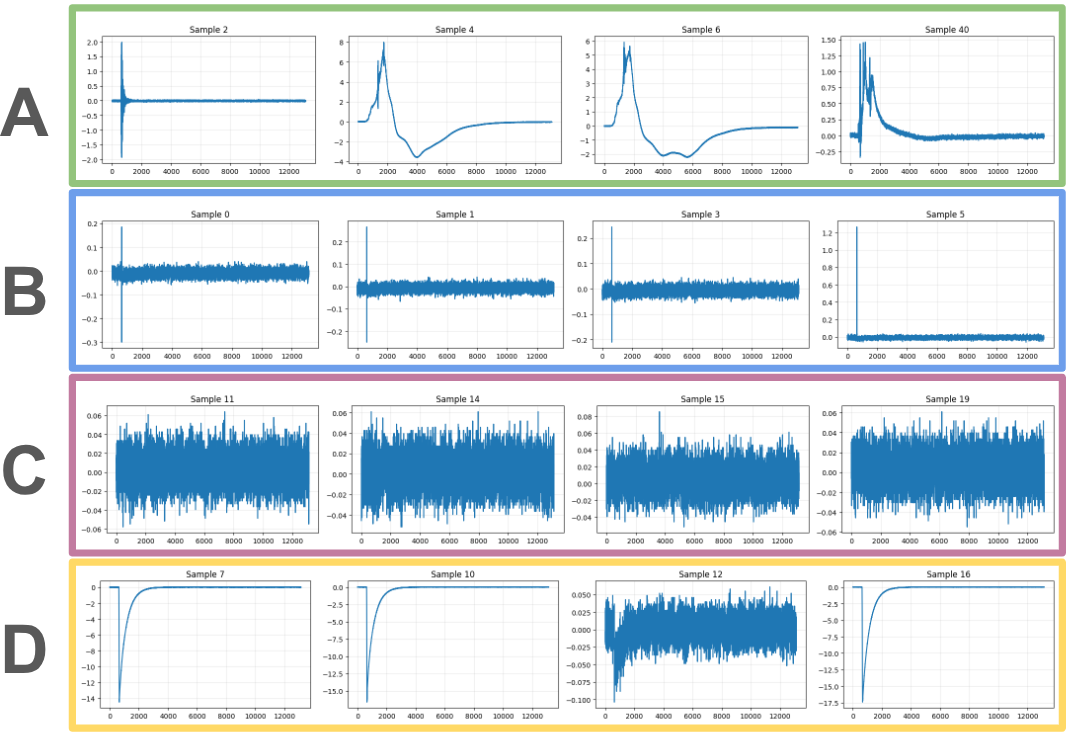}
    }
    \caption{Event classes identified by the UMAP classifier.} 
    \label{fig:classes}
\end{figure}
A set of classified signal segments is shown in Figure~\ref{fig:classes}. Class~A clearly contains the BD signals, whereas Classes~B and C contain triggered noise with possible digital spikes that are probably related to firmware issues and are still under investigation. Class~D reveals the false trigger events, clearly reproducing the profile shown in Figure~\ref{fig:fake_breakdown}.

We then applied this classification to improve the identification of the BD time in the TPS signals. This was implemented by tracking the progressive maximum derivative of both the potential and the current signals in order to identify the voltage drop associated with each BD. The result was finally matched to the absolute NTP time recorded by the THubs to identify the correct signal segments corresponding to the BD events.

\subsection{Signal Delays}
\label{sec:signal_delays}

Once all the BD segments were identified and the hubs timing corrected with the appropriate delays, we proceeded to determine the physical propagation delays of the signals along the MITICA transmission line for the different probes.
A selected dataset was then distilled by retaining only those pulses for which the signals recorded from both GP00 and the Rogowski coils belonged to Class~A, thus maximizing the likelihood of obtaining reliable signal matches.
During the test campaign the sampling frequency of the acquisition was lowered to allow the recording of longer trends, looking at oscillations lasting more than 100~$\mu s$, but such change resulted not suitable for fitting the relative shift. For this reason the final filtered dataset contains only 156 pulses.

An algorithm was implemented to determine the shift between two signals by maximizing the cross-correlation function; it can also optionally apply filtering and windowing, while constraining the maximum accepted correlation shift to a physically acceptable range. 

\begin{figure}
    \centering
    \includegraphics[width=0.45\textwidth]{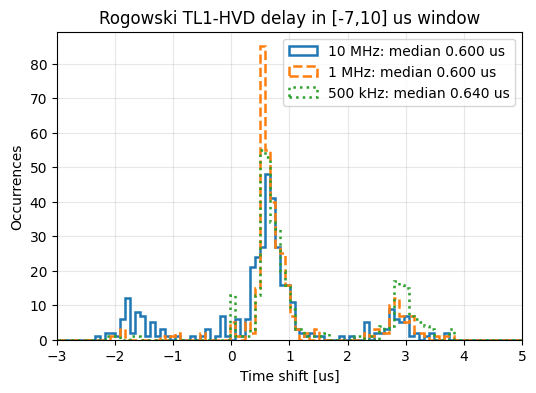}
    \caption{Histograms of the time delay between the HVD and TL1 Rogowski probe signals measured in Class~A pulses: (a) filtered at 10~MHz, (b) filtered at 1~MHz, and (c) filtered at 500~kHz.}
    \label{fig:rog_histograms}
\end{figure}

Figure~\ref{fig:rog_histograms} shows a histogram of the time delay between the signals measured by the two HVD and TL1 Rogowski probes for the Class~A pulses, with the signals filtered at 10~MHz, 1~MHz, and 500~kHz.
The result shows a consistent delay between the signals acquired by the two probes, with the HVD bushing signal leading the TL1 Rogowski signal by approximately 600~ns.
The same analysis was carried out to determine the delay between the GP00 Front, Center, and Rear signals, showing, as expected, delays of only a few nanoseconds: namely 20~ns between Front and Center, and 40~ns between Front and Rear.

The delay between the GP00 signals and the Rogowski set along the transmission line was less straightforward to determine because of the different waveform shapes caused by the complex circuit involved in closing the background current loop to ground. To improve the possibility of correlating these signals, we temporally aligned all the GP00 signals to the Front probe and the TL1 Rogowski to the HVD bushing signal.
\begin{figure}[h]
    \centering
    \subfigure[]{
        \includegraphics[width=0.48\textwidth]{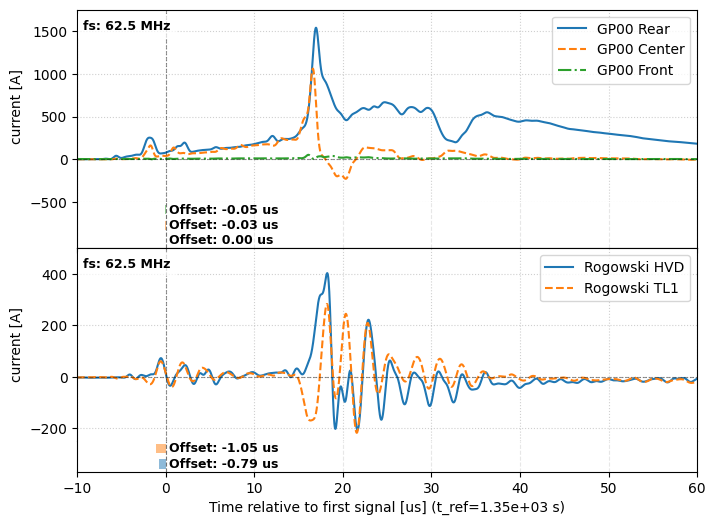}        
    }
    \subfigure[]{
        \includegraphics[width=0.48\textwidth]{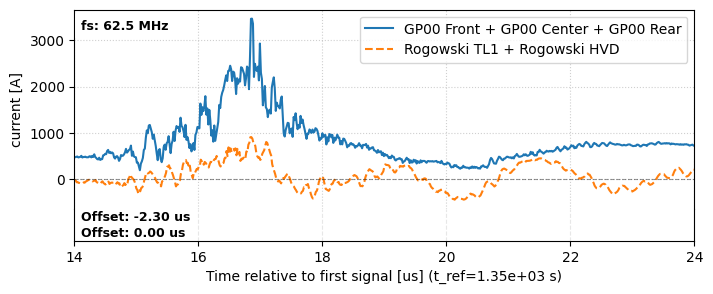}
        \label{fig:example_signal_sync_b}
    }
    \subfigure[]{
        \includegraphics[width=0.48\textwidth]{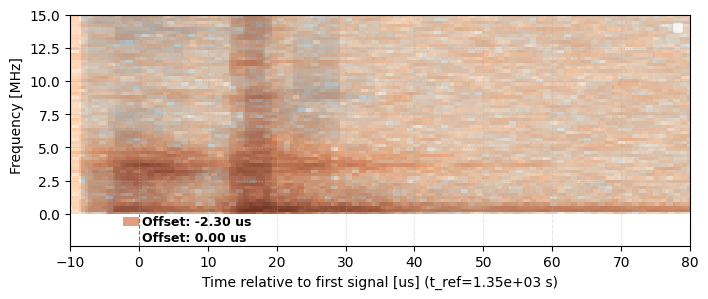}
    }
    \caption{Example of resynchronized signals for the event 2024072308. (a) top: GP00 signals filtered at 1~MHz and adjusted with 30~ns and 50~ns delays; bottom: Rogowski probes signals adjusted with 600~ns delay to align them with the HVD bushing signal position. (b) Sum of GP00 (solid-blue) and Rogowski probes (dashed-orange) in a zoomed window around the BD event [14,24]~$\mu$s, showing that after compensating for the 1.3~$\mu$s time delay the two signals are well aligned. (c) Spectrogram of the GP00 and Rogowsky probes, showing a 3.5~MHz resonance preceding and following the BD.} 

    \label{fig:example_signal_sync}
\end{figure}
An example of the resulting signals is shown in Figure~\ref{fig:example_signal_sync}. The first plot shows the three signals from GP00, filtered with a 4-pole zero-shift low-pass filter with 1~MHz bandwidth and corrected to match the timing of the Front plate. The bottom plot shows the current measured by the two Rogowski coils at the HVD bushing and at the TL1 middle point.

The overall correlation is weaker, as expected, because the current can follow different paths to close the circuit to ground during each BD event, including paths that return through the transmission line.
\begin{figure}
    \centering
    \includegraphics[width=0.45\textwidth]{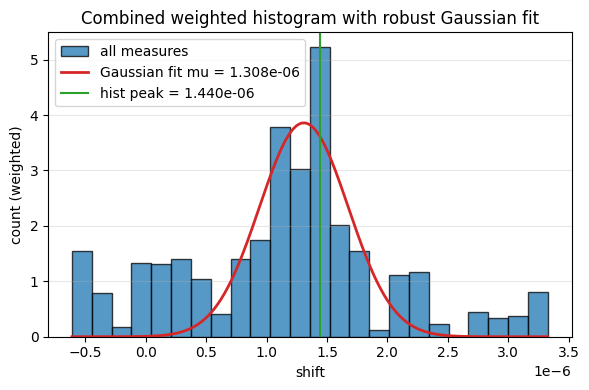}
    \caption{Histogram of the time delay between the GP00 and Rogowski probe signals measured in Class~A shots using the weighted cross-correlation method.}
    \label{fig:umap_ae2df}
\end{figure}
The plot shows a maximum correlation at around $1.3 \sim 1.4~\mu\text{s}$, which is not in perfect agreement with the expected delay.

\section{Discussion}
\label{sec:discussion}

The measured delay between the GP00 signals and the Rogowski signals appears larger than what would be expected from the sole propagation along TL2 and TL3. In first approximation, these two sections can be viewed as SF$_6$\dash insulated coaxial lines where the delay of propagation $\tau$ is driven by: 
\begin{equation}
\tau=\frac{\ell}{v_p}=\ell \cdot \sqrt{\mu\varepsilon}
\end{equation}
where $\varepsilon$ is the electric permittivity and $\mu$ is the magnetic permeability of the gas.
Assuming $\mu \approx \mu_0$ and $\varepsilon \approx \varepsilon_0$ for SF$_6$, for the 80~m length line, this results in $\tau \approx 80/c \approx 2.7\times10^{-7}$ s, namely about 270 ns. This value is much smaller than the observed $1.3\,\mu$s, confirming that the measured delay is not consistent with the transmission line alone.
A plausible explanation is that part of the additional lag is introduced by the inductance of the core-snubber protection, whose role in the fast transient dynamics is already known to be relevant~\cite{DAN2023113517,ZANOTTO2023113381}. 
If the hypothesis holds, we can roughly quantify the time component in terms of relative permeability of the core material:
\begin{equation}
    \tau_\text{cs} \simeq \ell_\text{cs} \cdot \sqrt{\frac{\mu_\text{cs}}{\mu_0}} \tau_0 
\end{equation}
where $\tau_0$ is the light propagation time on vacuum and $\mu_\text{cs}$ is the magnetic permeability of the core-snubber component.
Thus, for the core-snubber installed in TL3 having a length of $l_\text{cs} = 5 m$, with measured $\tau_\text{cs} \simeq 1.03 \mu s$, we obtain $\mu_\text{cs}/\mu_0\simeq4\cdot10^3$, that is almost compatible with the \textit{FINEMET~TF\dash1H} material vendor specs~\cite{FinemetTF1H}.
On the other hand, the discrepancy observed between the two Rogowski coils is more challenging to explain. It may be linked to the differing responses of the two sensors to an impulsive current, such as the event shown in Fig.~\ref{fig:example_signal_sync_b}. 
Such discrepancy likely arises from a combination of several phenomena, including: the transient impedance of the ground mesh during the BD current loop resulting in a current component also in the TL1 shell, the asymmetry introduced into this component by the core-snubber itself, and the capacitive coupling between the TL1 shell and the sensors.
A more precise assessment of the definitive delays among these components will require a detailed model and potentially additional diagnostic probes, and a dedicated investigation prior to the next experimental campaign. 
Nevertheless, even though the delay is larger than expected, once fully characterized it could provide a valuable diagnostic tool to improve the discrimination of future BDs occurring inside and outside of the BSV in MITICA.

\section{Conclusions}
\label{sec:conclusion}
This work presented the implementation of a general strategy for the accurate time synchronization of transient signals used in MITICA high-voltage holding tests.
The strategy involved repurposing available transient recorders to overcome components availability limitations and electrical constraints imposed by the harsh electromagnetic environment of the breakdown phenomena.
It was possible to obtain a consistent relative timing among the diagnostics and to analyze the propagation of the breakdown signals along the vessel and the transmission line.
The results highlighted a reproducible delay of about 600~ns between the two Rogowski probes and a larger delay of about 1.3~$\mu$s between the GP00 and the Rogowski signals.
Although these delays still require further investigation, the proposed approach demonstrates a practical and effective solution for the localization of breakdowns in MITICA and provides a basis for improving the discrimination of future out-of-vessel events.
Broadly speaking, this approach is particularly worth considering for experimental setups where the acquisition devices are not capable of direct synchronization while satisfying the assumptions given in Sec.~\ref{sec:redpitaya_rules}, providing an affordable, yet flexible and precise solution that can be implemented by leveraging a simple existing trigger distribution infrastructure.

\bibliographystyle{IEEEtran}
\bibliography{bibliography}

\end{document}